\documentclass[12pt,preprint]{aastex}

\slugcomment{submitted to ApJ}

\shorttitle{ULX in NGC 5408}
\shortauthors{Lang et al.}

\begin{document}

\title{A Radio Nebula Surrounding the Ultraluminous X-Ray Source in NGC 5408}

\author{Cornelia C.\ Lang, Philip Kaaret} 
\affil{Department of Physics and Astronomy}
\affil{University of Iowa,  Van Allen Hall, Iowa City, IA 52242, USA}
\email{cornelia-lang@uiowa.edu}
\author{St\'ephane Corbel} 
\affil{AIM$-$Unit\'e Mixte de Recherche CEA$-$CNRS}
\affil{Universit\'e Paris VII$-$UMR 7158, CEA Saclay, Service d'Astrophysique, F$-$491191 Gif sur Yvette, France.}
\and
\author{Allison Mercer}
\affil{Department of Physics and Astronomy}
\affil{University of Iowa,  Van Allen Hall, Iowa City, IA 52242, USA}

\begin{abstract}

New radio observations of the counterpart of the ultraluminous X-ray
source in NGC 5408 show for the first time that the radio emission is 
resolved with an angular size of $1.5\arcsec$ to $2.0\arcsec$. This
corresponds to a physical size of 35--46~pc, and rules out interpretation 
of the radio emission as beamed emission from a relativistic jet. In addition, the
radio spectral index of the counterpart is well determined from three
frequencies and found to be $\alpha$=$-$0.8 $\pm$0.2. The radio emission 
is likely to be optically-thin synchrotron emission
from a nebula surrounding the X-ray source.  The radio luminosity of
the counterpart is $3.8 \times 10^{34} \rm \, erg \, s^{-1}$ and the minimum energy
required to power the nebula is $\sim 1 \times 10^{49} \rm \, erg$.  These
values are two orders of magnitude larger than in any Galactic nebula powered
by an accreting compact object.

\end{abstract}

\keywords{black hole physics -- galaxies: individual: NGC 5408
galaxies: stellar content -- X-rays: galaxies -- X-rays: black holes}

\section{Introduction}

Ultraluminous x-ray sources (ULXs) are non-nuclear x-ray sources in
external galaxies with apparent luminosities above the Eddington limit
for a $20 M_{\odot}$ black hole, the maximum mass
of any dynamically-measured black hole mass in the Galaxy \citep{remillard06}; 
for a review, see \citet{fabbiano06}. 
ULXs with strong variability are likely accreting objects and may either be
``intermediate mass'' black holes
\citep{colbert99,makishima00,kaaret01} or normal (stellar-mass) black
holes with beamed or super Eddington radiation
\citep{king01,kording02,begelman02}.  The ULX in the dwarf irregular
galaxy NGC 5408 (NGC 5408 X-1) is variable and is one of the few ULXs
with a known radio counterpart \citep{kaaret03}.  The radio emission
could arise directly from a relativistic jet beamed toward our line of
sight, in which case a stellar-mass black hole would suffice to produce
the radio and X-ray emission, or from a nebula surrounding the compact
object. Recent radio observations presented by \citet{soria06} suggest
that the source has a steep spectrum ($\alpha$ $<$ $-$1), but show no 
indication of radio flux density variability. 

In order to better understand the nature of the radio emission from NGC
5408 X-1,  we obtained new joint observations in the radio using the Very
Large Array (VLA) of the National Radio Astronomy
Observatory\footnotemark\footnotetext{The National Radio Astronomy 
Observatory is a facility of the National Science Foundation operated
under cooperative agreement by Associated Universities, Inc.} and in 
the X-ray using the {\it Chandra X-Ray Observatory}. In addition, because
the target is a low-declination source ($\delta$ $\sim$ $-$41\arcdeg), we
present new and archival radio observations from the Australia Telescope Compact
Array (ATCA). Finally, we also reanalyzed archival optical
observations made using the {\it Hubble Space Telescope} (HST).  We describe
the observations and data reduction in \S~2,  present our results in
\S~3, and discuss the interpretation in \S~4.

\section{Observations and Analysis}

\subsection{VLA Observations}

VLA observations of the ULX source in NGC 5408 were made at 4.9~GHz
(6~cm) with four different array configurations between December 2003 and January 2005
: A, BnA, B, and CnB (see Table~\ref{obstable}). Each observation was approximately
3.5~hours long and corresponds to a joint {\it Chandra} X-ray 
observation. Standard procedures were carried out for flux and phase
calibration using the Astronomical Imaging Processing System (AIPS) 
of the NRAO. We used the flux density calibrator, J1331+305, and the phase
calibrator, J1353-441, a source which is located within $3\arcdeg$ of the target.  
For all observations, fast switching between the target source and
calibrator was used with a cycle time for the calibrator-source pair of
2.5 to 3 minutes (with 45 seconds on the calibrator and 90
to 120 seconds on the source).
  
The VLA image resolutions vary between $1.01\arcsec \times 0.28\arcsec$ (A-array)
and $6.01\arcsec \times 3.30\arcsec$ (CnB-array).  Because of the
proximity of the bright starburst region in NGC5408 and the weak radio emission
associated with the ULX, obtaining an accurate flux density for this
source is difficult.  However, with the high resolution of the VLA A, BnA and
B-arrays, it is possible to distinguish the radio emission associated
with the ULX from the extended emission from the starburst region in
most of our images.  

In fact, in attempting to determine the accurate flux density and
source structure, the extended emission from this starburst 
region turns out to be a key issue. One way to examine the contribution of the extended
emission is to adjust the imaging weighting function. The  Briggs'
ROBUST parameter controls the data weights in the {\it (u,v)} plane. 
Positive values of the ROBUST parameter (1 to 5) bring out the more
extended structure by weighting the inner part of the {\it (u,v)} plane
more heavily (``natural weighting''). Such values will increase the signal 
to noise on extended features but lower the spatial resolution. 
Negative values of the ROBUST parameter ($-$1 to $-$5) give more equal
weight to all points in the {\it (u,v)} plane (``uniform weighting'') which tends
to maximize spatial resolution but at the cost of signal to noise. 
A ROBUST=0 is balanced between these extremes. As many four images were made 
from data for an individual array using varying ROBUST parameters. 

In addition, a more quantitative way to remove extended flux is 
to limit the range of {\it (u,v)} baselines used for imaging. In
cases where the emission from the ULX is contaminated by extended
flux from the starburst, a {\it (u,v)} cutoff of 5 k$\lambda$ was 
applied. This cutoff limits flux on angular size scales larger
than $\sim$6\arcsec. For the VLA-B and VLA-CnB array images, where extended
flux is present, such cutoffs have been applied.

\subsection{ATCA Observations}

We also made continuum radio observations of NGC 5408 using the
Australia Telescope Compact Array (ATCA) located in Narrabri, New South
Wales, Australia.  The ATCA consists of five 22-m antennas positioned
along an E-W track with a short N-S spur and a sixth
antenna at a fixed location. We obtained three new ATCA observations 
of 12 hours each on consecutive days with the 6~km configuration 6.0A (6A). 
This provides the best angular resolution possible with ATCA ($\sim$1\arcsec~
at 4.8 GHz).  For the first two observations, we switched the feed-horns at roughly one hour
intervals observing in either the 6~cm band (4.8 GHz) or the
13 and 20~cm bands (1.4 and 2.4 GHz). The duty cycle of the
switching was arranged so the integration at 6~cm was four times as
long as those at  13 and 20~cm.  On the last day, we observed with the 
5 cm band (6.1 GHz) instead of the 6 cm band (4.8 GHz). 

J1934-638 and J1349-439 were used for flux and phase calibration.
Editing, calibration, imaging and analysis were performed using 
standard routines in the MIRIAD software package
\citep{sault98}. A radio counterpart to the ULX was detected at
all frequencies.  Because of the low declination of NGC 5408, the
ATCA forms a more symmetric synthesized beam than the VLA. The
extended emission associated with the starburst in NGC 5408 is present,
but in the ATCA images, the radio counterpart to the ULX is 
clearly distinguished, especially in the 6A configuration
data at 4.8, 6.2 GHz. However, at the lower frequencies (1.4 and 2.3 GHz) 
there is still contamination from the starburst. During imaging, a variety 
of ROBUST parameters were also used to
downweight the extended emission, similar to the procedure in the 
VLA imaging. In addition, we analyzed 
archival ATCA observations 
of NGC 5408 at a range of frequencies, observed during 2000-2004 
(listed in Table~\ref{atca_obs}). 
In the several of the ATCA datasets, (e.g., the 6D-array 4.9 GHz
data (Kaaret et al. (2003), and the 2.3 and 1.4 GHz archival datasets)  
a {\it (u,v)} cutoff range of 3 k$\lambda$ was used to remove flux on 
extended scales larger than 11\arcsec. 

\subsection{X-ray Observations}

NGC 5408 was observed with the {\it Chandra X-Ray Observatory}
(Weisskopf 1988) five times using the ACIS spectroscopic array (ACIS-S;
Bautz et al.\ 1998) in imaging mode and the High-Resolution Mirror
Assembly (HRMA; van Speybroeck et al.\ 1997). The observations began in
2002 and ended in 2005, see Table~\ref{obstable}.  Because a high X-ray
count rate was expected,  only the S3 chip was operated with a 1/4
sub-array mode for the first observation and a 1/8 sub-array for the
other observations.  This gave an exposure time of 0.8~s for the first
observation and 0.4~s for the others.  For each observation, we constructed 
an image using all valid events on
the S3 chip and used the {\it wavdetect} tool which is part of the {\it
CIAO} data analysis package to search for X-ray sources.  Typically
only the ULX was detected in each observation.   The position of the
ULX was within $0.23\arcsec$ of RA = 14h03m19.63s $\pm$ 0.01s, DEC
-41d22'58.65" $\pm 0.2\arcsec$ (J2000) in all observations.

We fit the X-ray spectrum of the ULX for each observation using the
{\it XSPEC} spectral fitting tool which is part of the {\it LHEASOFT}
X-ray data analysis package and response matrices calculated using {\it
CIAO}.  As previously reported by \citet{kaaret03}, we found that
absorbed single power-law models did not provide an adequate fits
(except for the last observation), while models consisting of either an
absorbed broken power-law or the absorbed sum of a multicolor disk
blackbody plus a power-law did provide adequate fits.  There were no
pronounced changes in spectra shape between the different
observations.  The fit parameters were similar to those quoted in
\citet{kaaret03}.  The flux for each observation calculated using the
fitted multicolor disk blackbody plus power-law model are given in
Table~\ref{obstable}. We also did not observe any significant changes
in the X-ray flux level between observations.

\subsection{HST Observations}

Two WFPC2 observations of NGC 5408 are present in the {\it Hubble Space
Telescope} (HST) archive.  The observations were obtained as part of a
snapshot survey of nearby dwarf galaxy candidates (GO-8601, PI Seitzer)
and consist of 600~s WFPC2 exposures with the F606W and F814W filters
obtained on 4 July 2000.  After extracting the data from the HST
archive, we used IRAF to mosaic and clean the images.  We corrected the
absolute astrometry of the HST image using stars from the 2MASS catalog
\citep{2mass} using the Graphical Astronomy and Image Analysis Tool
(GAIA).  We used only 2MASS sources on the WF3 chips where the ULX is
located in order to preclude issues regarding the relative positioning
of the WFPC2 chips.  We estimate that the astrometric uncertainty is
$0.2\arcsec$.  We used the HSTphot stellar photometry package to obtain
photometry \citep{dolphin00}.  We removed bad pixels, cosmic rays, and
hot pixels and then obtained simultaneous photometry for the F606W and
F814W images.

\section{Results}

\subsection{Radio Counterpart}

The flux density of the radio emission associated with the ULX in
the VLA and ATCA images was measured by fitting a 2-D Gaussian
point source to the peak radio emission in each of the various
images. In all cases, the source appears to be point-like with
slightly extended emission, but nothing which we can confidently 
resolve.  The radio source position 
from our VLA B-array image (resolution of
$3.17\arcsec \times 0.92\arcsec$) is RA, DEC (J2000) = 14h03m19.63s 
$\pm$ 0.01s, $-$41d22\arcmin~58.7\arcsec~ $\pm 0.2\arcsec$, which
is within $0.1\arcsec$ of the ULX position
calculated from the {\it Chandra} observations \citep{kaaret03}.  This is
well within the {\it Chandra} error circle. Using the ATCA 4.8~GHz and
6.1~GHz data, the position of the radio source is RA, DEC (J2000) = 14h03m19.61s
$\pm$ 0.02s, $-$41d22'58.5" $\pm$0.2".

\subsubsection{4.9 GHz Emission}

\begin{figure}
\centerline{\includegraphics[angle=270,width=6in]{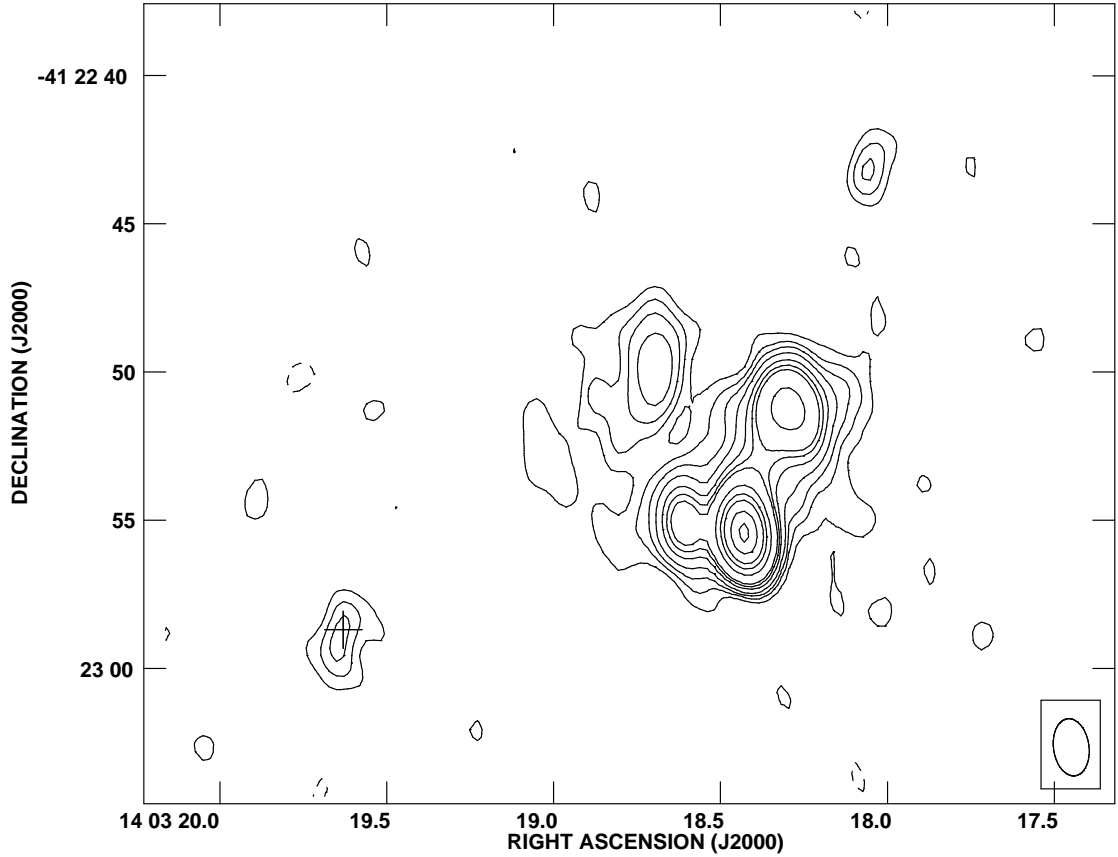}}
\caption{VLA 4.9 GHz BnA-array image of the radio emission in NGC5408.
The compact radio source associated with the ULX is located in the
lower left (SE) of the plot. The cross represents the position of the
ULX. The spatial resolution of this image
is 1.94\arcsec~$\times$ 1.20\arcsec, PA=8\arcdeg, and the image has been made with
ROBUST=1 weighting (slightly naturally weighted). The contour levels represent 
radio emission at -3, 3, 5, 7, 9,  10, 15, 20, 25, 50, 75, 100, 150, and 200 times the rms
level of 20 $\mu$Jy beam$^{-1}$.} 
\label{vla_image} 
\end{figure}

Table 3 lists the measured 4.9 GHz flux density and corresponding 
ROBUST weighting, array configuration, geometrical beam size 
($\sqrt{\theta_{maj}\theta_{min}}$) and an indication of any
{\it (u,v)} cutoff used in the imaging.
Radio emission was detected above 3$\sigma$ in 20 of the 21 images
made at 4.9 GHz from the VLA and ATCA data. 
The source was not detected in the highest resolution 4.9 GHz VLA image 
(1.01\arcsec~$\times$ 0.28\arcsec; A-array, Robust=$-$1) with an upper 
limit of 0.1 mJy. The largest angular size detectable with the VLA in
its A-array configuration is 10\arcsec. The radio source is detected in 
the other two VLA A-array images, but with modest detections 
($\sim 7\sigma$ in both cases). Figure~\ref{vla_image} shows the BnA-array image of
NGC5408 (Robust=1). The resolution of this image is $1.94\arcsec \times
1.20\arcsec$ and the source is clearly detected at the $10\sigma$ level. 

\begin{figure}
\centerline{\includegraphics[angle=270,width=6in]{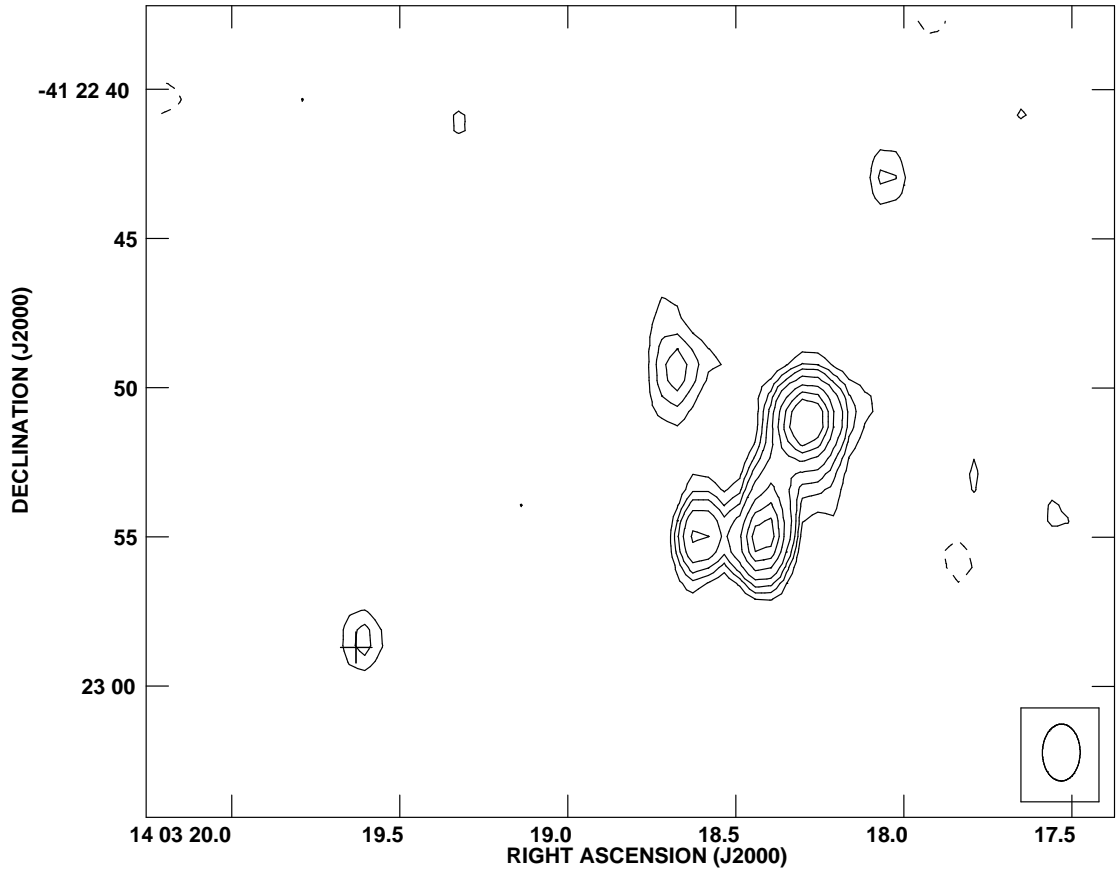}}
\caption{ATCA 4.8 GHz image of the radio emission in NGC 5408. The 
compact radio source associated with the ULX is located in the
lower left (SE) of the plot. The crosses represents the position of the 
ULX. The spatial resolution of this image
is 1.90\arcsec~$\times$ 1.26\arcsec, PA=$-$55\arcdeg, and contours
represent radio emission at $-$3, 3, 4, 5, 7, 9, 11, 13, 18, 21, 25, 30, 52, 84, 
and 136 the rms level of 32 $\mu$Jy beam$^{-1}$. The image was made
with uniform weighting (ROBUST=-5). 
\label{atca_unif}} \end{figure}

\begin{figure}
\centerline{\includegraphics[angle=0,width=6in]{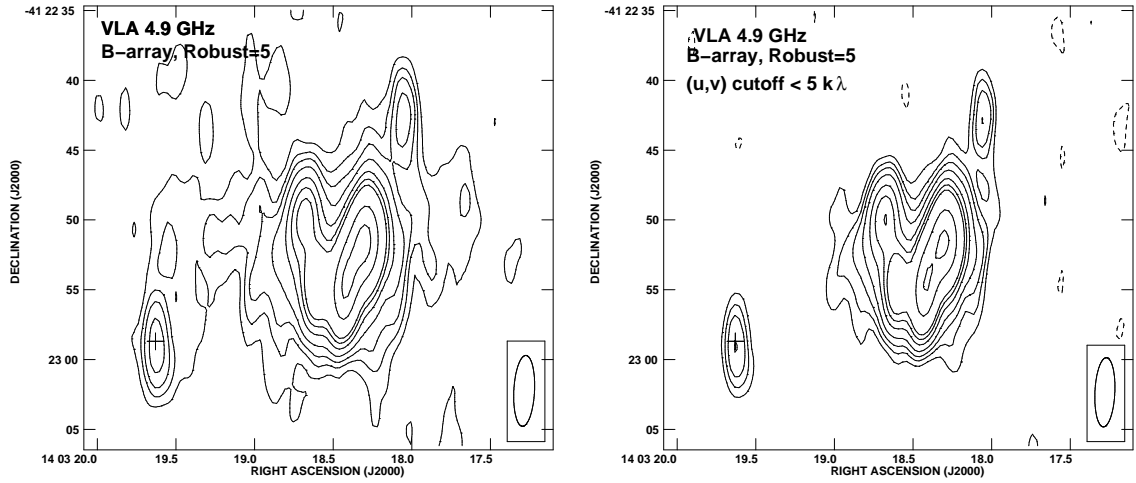}}
\caption{VLA 4.9 GHz B-array images of the radio emission in NGC 5408
made with ROBUST=5 (natural weighting). The left panel shows an image with 
a spatial resolution and the right panel shows an image made with data
which have had a ({\it u,v}) cutoff $<$ 5 k$\lambda$ applied to them.  
In both cases, the spatial resolution is 5.0\arcsec~$\times$1.5\arcsec, 
PA=$-$3\arcdeg, and contour levels reprsent $-$3, 3, 5, 7, 10, 15, 20, 25, 
50, 75, 100, 125 and 150 times the rms level of 20 $\mu$Jy beam$^{-1}$. 
\label{vla_barray}} \end{figure}

Figure~\ref{atca_unif} shows the highest resolution 4.9 GHz ATCA image,
which has a beamsize very similar to the VLA BnA-array image shown in 
 Figure~\ref{vla_image}. In the majority of the VLA and ATCA images, the 
radio emission associated with NGC 5408 X-1 is obvious and clearly separated from
the extended emission associated with the NGC 5408 starburst. However, in cases 
such as the VLA B-array naturally-weighted image, where the resolution is lower and
favors the extended structures, a significant amount of extended emission is 
present (see Fig.~\ref{vla_barray}, left). Removing the shortest 
({\it u,v}) baselines ($<$ 5 k$\lambda$) for these lower-resolution images 
produces an image, shown in Fig.~\ref{vla_barray}, right, where the the radio 
emission associated with NGC 5408 X-1 is clearly
separated from the extended emission associated with the starburst.

\begin{figure}
\centerline{\includegraphics[angle=90,width=6in]{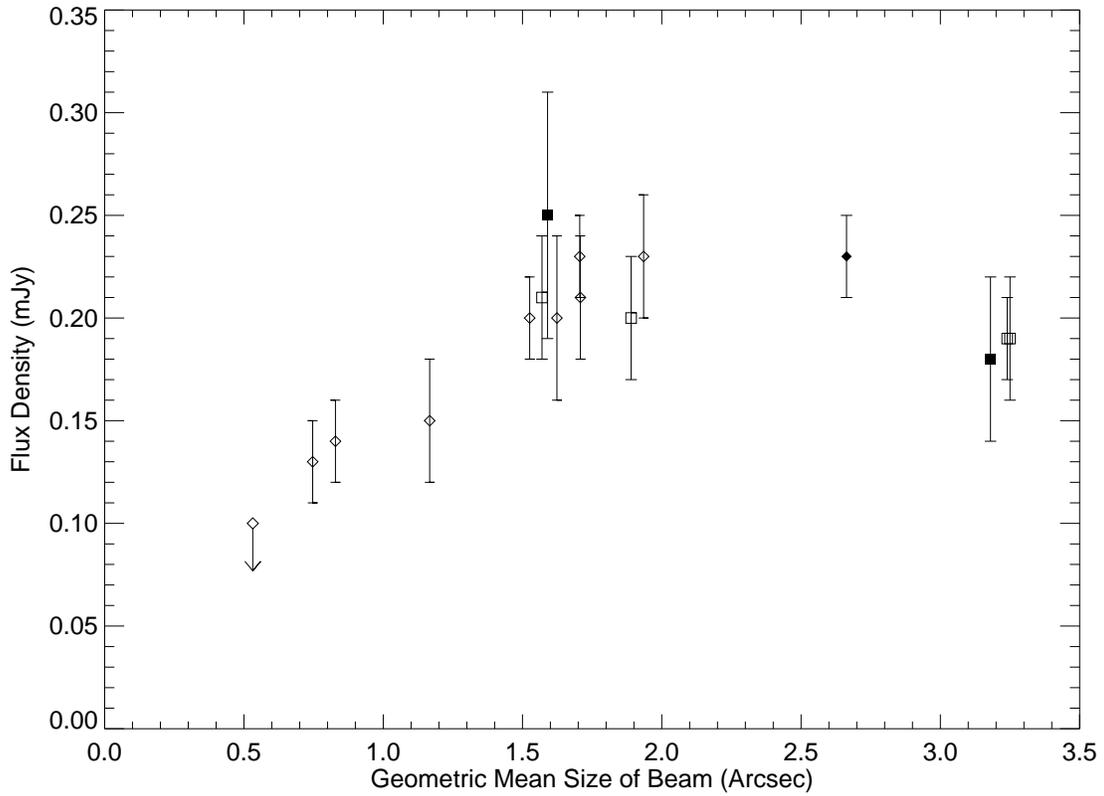}}
\caption{Flux Density vs. Geometric Beamsize for ATCA (squares) and VLA 
(diamond) radio observations at 4.9 GHz from this paper and from
archival observations of Soria et al. (2006). Filled symbols represent 
measurements from images where the shortest ({\it u,v}) baselines
have been removed to reduce contamination from the background galaxy. 
Error bars represent the error in fitting 2-D Gaussians to the
unresolved point source. 
\label{flux_vs_beam}} 
\end{figure}

Figure~\ref{flux_vs_beam} shows the flux density and associated errors
versus the geometric beamsize for measurements from 16 images of
VLA (diamonds) and ATCA (square; new and archival) data taken from 
Table~\ref{ulx_radioflux}. The filled 
symbols indicate that a {\it (u,v)} cutoff was applied to the data in order
to remove extended emission which may contaminate the flux density. 
However, not all measurements were included in Figure~\ref{flux_vs_beam}. 
Images made with the VLA CnB-array at all robust weightings and with
the ATCA 6D configuration have low enough resolution that it is difficult
to separate the background emission from NGC 5408 from the radio emission
at the position of the ULX. In fact, the flux densities for the VLA CnB-array 
and ATCA 6D-array measurements are in the range of 0.32 to 0.51 mJy, up to twice as 
high as measurements made
with more extended arrays, indicating that the background contributes
significantly to the measurement. For this reason, only the measurements from
images where we were confident we were separating out the radio emission
associated with NGC 5408 X-1 were used to make Figure~\ref{flux_vs_beam}. 

\subsubsection{Source Size and Structure}

Previously, \citet{soria06} had made some of the highest resolution
observations of the ULX in NGC 5408 using the ATCA. They measured 
the flux density in images with beamsizes of $1.5\arcsec$ 
to $3.5\arcsec$ and found an essentially flat distribution of 
flux density with beamsize. However, their observations do not resolve
the source and therefore are only consistent with emission from 
a point source (Soria et al. 2006). The VLA data presented here 
for the first time probe even higher angular resolutions (0.5\arcsec~to 1.5\arcsec). 
\ref{flux_vs_beam} shows that the 
VLA data are crucial in studying the flux density versus
geometric beam size for scales $<$ 1.5\arcsec. There is a clear 
trend of decreasing flux density at smaller
beamsizes below a geometric beamsize of 1.5\arcsec.  
The linear rise of flux density with beamsize shown in
Figure~\ref{flux_vs_beam} indicates that that there is extended
emission associated with this source on beamsize scales between $\sim
0.5\arcsec$ and $\sim1.5\arcsec$. For beamsizes greater than $\sim$ 
$1.5\arcsec$, the flux density measurements are relatively constant
(within the errors), suggesting the source has an angular extent in
the range of $1.5\arcsec$ to $2.0\arcsec$. 

\subsubsection{Variability}

Because the radio source associated with the ULX is likely to be somewhat
extended, different configurations of the VLA radio telescope will be 
sensitive to emission on differing angular size scales. Therefore, it is 
not possible to look for flux variability in our high-resolution VLA observations (where the 
configuration ranges from A-array to CnB-array). However, ATCA observations 
over a number of epochs (e.g., \citet{kaaret03}, \citet{soria06}, and this paper) have
shown that the flux density at 4.9 GHz does not appear to vary significantly
and has an average, overall value near 0.20 mJy.

\subsubsection{Multifrequency Data and Spectral Index}

\begin{figure}
\centerline{\includegraphics[angle=0]{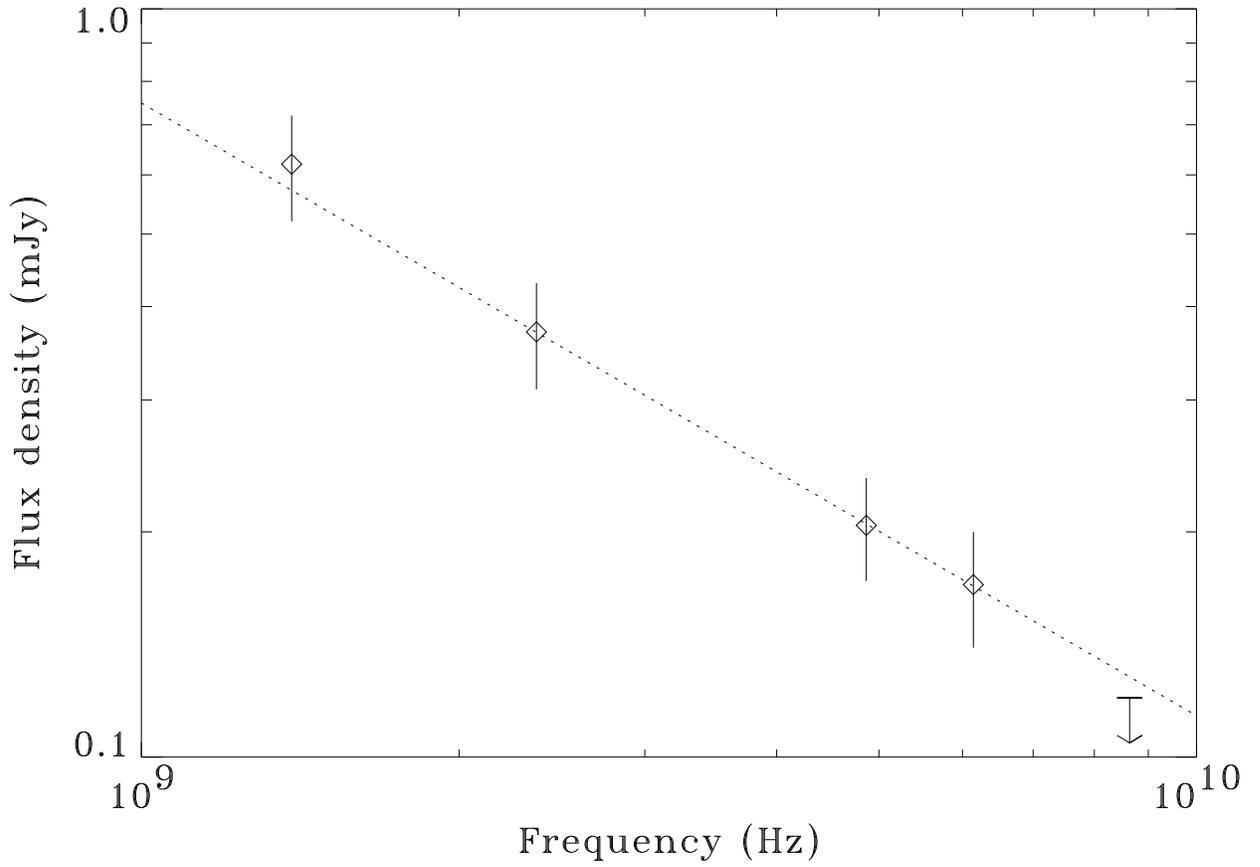}}
\caption{Flux density vs. frequency for the radio emission associated 
with NGC 5408 X-1. These measurements are from ATCA data, and the 
spectral index of $\alpha$=$-$0.82$\pm$24 is based on the 2.3, 4.9 and 
6.1 GHz points. The 1.4 GHz flux density and 8.5 GHz upper limit 
are also included on this plot. 
\label{spectrum_3pts}} 
\end{figure}

Multi-frequency observations at 1.4, 2.3, 4.9 and 6.1 GHz 
were made as part of the new ATCA data presented here. At each frequency, 
all existing ATCA data were combined (see
Table 2) and a point source was fit to the radio emission associated
with the ULX. We find flux densities of $0.62 \pm$ 0.10 mJy at 1.4 GHz 
(geometric beam size of 6.7\arcsec; ROBUST=$-5$; ({\it u,v}) cutoff 
$<$ 5 k$\lambda$), $0.37 \pm 0.08$ mJy at 2.4~GHz (geometric beam size 
of 3.3\arcsec; ROBUST=$-5$), $0.20 \pm 0.03$ mJy at 4.9~GHz (geometric 
beam size=1.5\arcsec, ROBUST=$-$5) and $0.17 \pm 0.03$ at 6.1~GHz.  
At 1.4 GHz, the contribution from the extended background
may be present even though we used uniform weighting and
limited the shortest {\it (u,v)} data; the beamsize is large. 
At 2.3 GHz, the image is made with uniform weighting and the
contribution from the extended starburst region is not apparent. 
 Therefore, we determine the spectral index based on the 
2.4, 4.9 and 6.1 GHz measurements only and
obtain $\alpha$=$-$0.8 $\pm$0.2, where S$_{\nu}$ $\propto$ $\nu^{\alpha}$. 
Figure~\ref{spectrum_3pts} shows the flux density versus frequency and includes the
1.4 GHz measurement as well as the 8.5 GHz ATCA upper limit from Kaaret et al. (2003). 
Although not included in the fit, the 1.4 and 8.5 GHz 
measurements are consistent with the fitted spectral index. 

\subsection{Optical counterpart}

\begin{figure}
\centerline{\includegraphics[angle=0,width=4in]{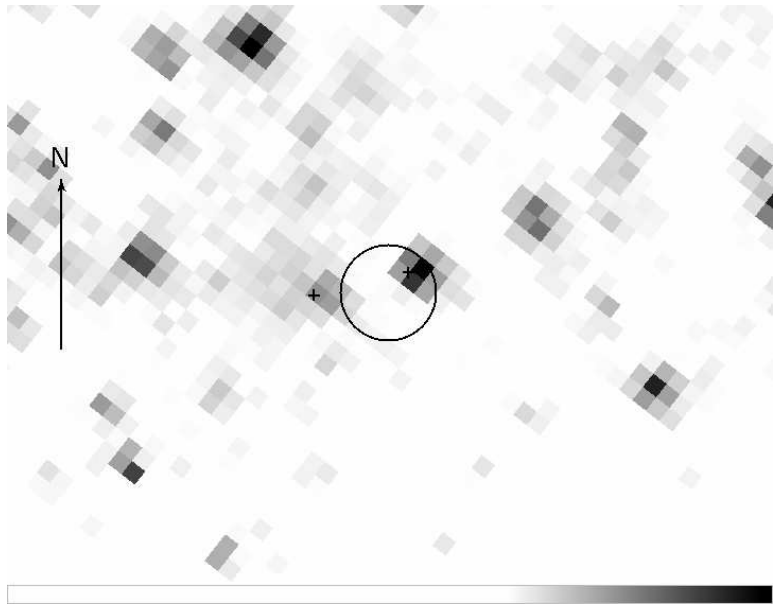}}
\caption{Image of the field near NGC5408 X-1 in the F606W band.  The
circle is centered at the position of the VLA counterpart of the ULX
and has a radius of $0.28\arcsec$, which represents the relative
position uncertainty including both the radio position uncertain and
the optical astrometry uncertainty. Only one source lies within the
error circle and we identify it as the likely optical counterpart of
the ULX.  The arrow has a length of $1\arcsec$ and points north.
\label{ulxclose}} \end{figure}

Figure~\ref{ulxclose} shows the HST/WFPC2 image in the F606W filter for
the area near the ULX.  The circle drawn on the figure is centered at
the VLA source position quoted above and has a radius of $0.28\arcsec$,
which represents the relative position uncertainty including both the
radio position uncertain and the optical astrometry uncertainty.  Only
one object lies within the error circle. It is located at a position of
RA, DEC (J2000) = 14h03m19.62s, $-$41d22'58.54 (J2000) which is $0.17\arcsec$
from the radio position and well inside the error circle. The next
closest HST source is $0.43\arcsec$ from the VLA position and is
outside the error circle.  We identify the HST source within the error
circle as the likely optical counterpart of the ULX.

The optical counterpart has magnitudes, in the WFPC2 flight photometric
system, of $22.387 \pm 0.021$ in the F606W filter and $22.396 \pm
0.043$ in the F814W filter (Dolphin 2000).  The equivalent Johnson-Cousins 
magnitudes are $V = 22.4$ and $I = 22.4$.  We corrected for reddening using an extinction
$\rm E(B-V) = 0.068$ based on the dust maps of \citet{schlegel98} and
using an $\rm R_V = 3.1$ extinction curve.  The dereddened magnitudes
are $V_0 = 22.2$ and $I_0 = 22.3$ and the color is $V_0 - I_0 = -0.1
\pm 0.1$.

\section{Discussion}

The radio counterpart size of $1.5\arcsec$ to $2.0\arcsec$ corresponds
to a physical diameter of 35--46~pc at the distance of 4.8~Mpc to NGC
5408, as determined by from the tip of the red giant branch
\citet{karachentsev02}.  Therefore, we can rule out the interpretation
of the radio emission associated with the NGC 5408 ULX as a
relativistically beamed jet \citep{kaaret03}.  

The radio surface brightness of NGC 5408 at 4.9 GHz is 
$\sim$4 $\times$ 10$^{-20}$ W m$^{-2}$ Hz$^{-1}$ sterrad$^{-1}$, 
assuming a source size of 1.5\arcsec. Taking a spectral index
of $-$0.8, we can calculate the surface brightness at 1.4 GHz
in order to compare to values for known supernova remnants
of similar sizes \citep{green04}. The value of 
$\sim$1 $\times$ 10$^{-19}$ W m$^{-2}$ Hz$^{-1}$ sterrad$^{-1}$
is significantly higher than that of known supernova
remnants of similar size. It is therefore more likely that 
the radio emission associated with NGC 5408 X-1 arises 
instead from an extended radio lobe. The radio spectrum 
($\alpha$=$-$0.8) of this source is consistent with 
optically-thin synchrotron emission.

For comparison, the radio nebula W50 surrounds the relativistic jet source 
SS 433 \citep{margon84} and is thought to be powered by the relativistic
outflow.  SS 433 has been suggested as a possible Galactic analog to
the ULXs \citep{fabrika01,begelman06}. The radio
nebula in W50 is interesting because the nebular radio emission is unlikely to
be beamed.  The physical size of W50 is roughly 50~pc and is similar to
what we estimate for the radio nebula surrounding NGC 5408 X-1.  The
radio spectral indexes of the various components of W50 range from
$-0.5$ to $-0.8$, also consistent with the spectral index of the radio
emission in NGC 5408 X-1 ($\alpha$ $\sim$$-$0.8).  A comparison of the radio 
brightness can be made for the two sources. W50 has an integrated flux density 
at 1.4 GHz of $\sim$ 70 Jy and a distance of $\sim$5 kpc (Dubner
et al. 1998). Assuming a spectral index of $-$0.5, that translates
to a 4.9 GHz flux density of 40 Jy, and if it were at the distance of
NGC 5408 (4.8 Mpc), its flux density would be $\sim$ 40 $\mu$Jy. 
The integrated flux density of NGC 5408 X-1 at 4.9 GHz is $\sim$200 $\mu$Jy, 
so its radio brightness is more than a factor of 5 greater than that of W50. 

We investigate the energetics of the radio lobe assuming radiation via
synchrotron emission, equipartition between particles and fields, and
equal energy in electrons and baryons.  We use a spectral index $\alpha
= -0.8$ (see $\S$3.1.4), a lower frequency cutoff of 1.3~GHz,
and an upper frequency cutoff of 6.2~GHz.  The total radio luminosity
of the source is $3.8 \times 10^{34} \rm \, erg \, s^{-1}$. For a
source diameter of 46~pc and a filling factor of unity, we find that
the total energy required is $3.6 \times 10^{49} \rm \, erg$, the
magnetic field is $16 \mu$G, and the synchrotron lifetime is $\sim$20~Myr. 
For contrast, for a diameter of 35~pc and a filling factor of 0.1, the total energy
required is $9 \times 10^{48} \rm \, erg$, the magnetic field is $39
\mu$G, and the lifetime is $\sim$5~Myr.  We note that the estimates of
\citet{soria06} for the energy content of a synchrotron nebula
surrounding NGC 5408 X-1 are about an order of magnitude larger.  This
is primarily because they assume an electron energy
distribution that extends down to a Lorentz factor of 1, while we,
conservatively, assume that the electron energy distribution extends
only over the range needed to produce the observed radio emission (1.3 to 6.2 GHz).

The total energy content in relativistic electrons in W50 is in the range
$(0.5-7) \times 10^{46} \rm \, erg$ \citep{dubner98}.  The total energy
content in relativistic electrons in the radio nebula surrounding NGC
5408 X-1 is at least two orders of magnitude larger.  Therefore, if two
nebula are similar, then the jet powering the nebula surrounding NGC
5408 X-1 must be at least two orders of magnitude more powerful than
that from SS 433 powering W50.  The radio lobes powered by the
persistent accreting stellar-mass black hole GRS 1758-258 have a radio
luminosity of $3 \times 10^{30} \rm \, erg \, s^{-1}$ and require an
energy content of $2 \times 10^{45} \rm \, erg$ \citep{rodriguez92} and
are even less powerful than W50.  We note that the ULX Holmberg II X-1
has an associated radio nebula with a total energy similar to that of
the NGC 5408 X-1 nebula \citep{miller05}.

At the distance to NGC 5408, the absolute magnitude of the optical
counterpart would be $M_V = -6.2$.  If the light arises only from the
stellar companion, then the magnitude and color exclude main sequence
and giant stars and require a supergiant star.  The absolute magnitude
and color are consistent with classification as a B or early A
supergiant.  The luminosity of such a star would be in the range of 1$-$5 $\times
10^{38} \rm \, erg \, s^{-1}$.  However, the stellar classification is
suspect since light may arise from reprocessing of X-rays from the
compact object and since the X-ray luminosity ($\sim 10^{40} \rm \, erg
\, s^{-1}$ if the emission is isotropic) exceeds the expected stellar
luminosity by a factor of at least 20 and may strongly affect the
physical state of the star.  

The X-ray to optical flux ratio, defined following \citet{vp95} as $\xi
= B_{0} + 2.5 \log F_{X}$ where $F_{X}$ is the X-ray flux density at
2~keV in $\mu$Jy and we approximate $B_0 = V_0$ due to lack of a B-band
image, is $\xi = 20.0$.  If we use the flux density at 1~keV, this
rises to $\xi = 21.1$.  These values are higher than those of any
high-mass X-ray binary (HMXB), other than LMC X-3, and are in the range
typically found for low-mass X-ray binaries (LMXBs).  Thus, it is possible 
that the companion star contributes little to 
the observed optical emission, e.g., it is not supergiant, and that most of
the optical light arises from reprocessing of X-rays, as occurs in LMXBs 
\citep{kaaret05}.

\section*{Acknowledgments}

PK acknowledges partial support from {\it Chandra} grant CXC GO4-5035A and a
Faculty Scholar Award from the University of Iowa.  STSDAS and PyRAF
are products of the Space Telescope Science Institute, which is
operated by AURA for NASA. The authors also thank Michael Rupen for planning and 
carrying out the VLA observations. 


\clearpage

\begin{deluxetable}{ccl} \tabletypesize{\scriptsize}
\tablecaption{Chandra and VLA Observations of NGC 5408 \label{obstable}}
\tablewidth{0pt} 
\tablehead{\colhead{Date/Time} & \colhead{X-Ray Flux} & \colhead{VLA array}}
\startdata 
2002-05-07 13:45:47 & 1.42 & $-$\\ 
2003-12-20 13:06:33 & 1.49 & B \\
2004-02-09 10:29:47 & 1.58 & CnB \\ 
2004-12-20 13:14:36 & 1.47 & A \\ 
2005-01-29 11:02:49 & 1.50 & BnA \\ 
\enddata 
\vspace{-12pt}\tablecomments{The table lists the Chandra/VLA
observations of NGC 5408 and includes the date and UT time of the start
of the Chandra observation (the VLA observations were roughly
simultaneous),  the ULX X-ray flux in the 0.3-8~keV band in units of
$10^{-12} \rm \, erg \, cm^{-2} \, s^{-1}$ during the observation, and
the VLA array configuration.} 
\end{deluxetable}

\clearpage
\begin{deluxetable}{lccl} \tabletypesize{\scriptsize}
\tablecaption{ATCA Observations of NGC 5408 \label{atca_obs}}
\tablewidth{0pt} 
\tablehead{\colhead{Date/Time} & \colhead{Array} & \colhead{Frequency (GHz)} & \colhead{Reference}}
\startdata 
2000-04-01 02:09&6D & 4.8, 8.6 & Kaaret et. al (2003)\\
2003-03-03 05:31&6A& 4.8, 8.6 & Archival 1 \\
2003-05-20 22:05&1.5C& 1.4 & Archival 2 \\	 	 
2003-07-27 17:45&6D& 1.4 & Archival 3\\	
2003-12-07 10:05&6A& 2.3, 4.8 & this paper\\
2003-12-08 11:30&6A& 2.3, 4.8 & this paper\\
2003-12-09 11:02&6A& 2.3, 6.1 & this paper\\
2004-12-09 to 12-11&6D& 1.4, 2.3 & Soria et al. (2006)\\
2004-12-09 to 12-11&6D& 4.8, 6.1 & Soria et al. (2006)\\
\enddata 
\vspace{-12pt}\tablecomments{The table lists the ATCA observations of
NGC 5408 and includes the observation date, the frequency
or frequencies of observation, and the measured flux density.}
\end{deluxetable}

\clearpage
\begin{deluxetable}{lrccc} \tabletypesize{\scriptsize}
\tablecaption{4.9 GHz Flux Density of NGC 5408 \label{ulx_radioflux}}
\tablewidth{0pt} 
\tablehead{\colhead{Telescope} & \colhead{Robust} & \colhead{Flux Density} & \colhead{Geometric} &\colhead{(u,v)}\\
\colhead{Array} & \colhead{Weight}& \colhead{(mJy)} & \colhead{Beamsize} &\colhead{cutoff}}
\startdata 
VLA-A&$-$1&$<$0.10$\pm$0.03&0.53&\\
VLA-A&1&0.13$\pm$0.02&0.75&\\
VLA-A&3&0.14$\pm$0.02&0.83&\\
VLA-BnA&$-$1&0.15$\pm$0.03&1.17&\\
VLA-BnA&1&0.20$\pm$0.02&1.53&\\
VLA-BnA&3&0.23$\pm$0.02&1.70&\\
VLA-B&$-$1&0.21$\pm$0.03&1.71&\\
VLA-B&0&0.23$\pm$0.03&1.93&\\
VLA-B&$-$5&0.20$\pm$0.03&1.62\\
VLA-B&5&0.23$\pm$0.02&2.66&$<$ 5 k$\lambda$\\
ATCA-6A&$-$5&0.21$\pm$0.03&1.57&\\
ATCA-6A&0&0.20$\pm$0.03&1.89\\
ATCA-6A&2&0.19$\pm$0.02&3.25\\
ATCA-6A&5&0.19$\pm$0.02&3.24\\
ATCA-6D&$-$5&0.25$\pm$0.06&1.59&$<$ 3 k$\lambda$\\
ATCA-6D&5&0.18$\pm$0.04&3.18&$<$ 3 k$\lambda$\\
\hline
VLA-CnB&0&0.51$\pm$0.03&4.45&\\
VLA-CnB&$-$5&0.46$\pm$0.04&3.77\\
VLA-CnB&$-$5&0.32$\pm$0.04&3.71&$<$ 5 k$\lambda$\\
VLA-CnB&$-$1&0.47$\pm$0.03&4.08\\
ATCA-6D&5&0.43$\pm$0.04&3.46\\

\enddata 
\vspace{-12pt}\tablecomments{The table lists the 4.9 GHz flux
densities for the radio counterpart of the ULX in NGC5408. 
All fits are for an unresolved point source. It is not possible
to distinguish between the source and background emission for
the measurements listed below the horizontal line.}
\end{deluxetable}

\clearpage

\clearpage

\clearpage

\clearpage

\clearpage

\clearpage


\begin{thebibliography}{}


\bibitem[Arnaud(1996)]{1996ASPC..101...17A} Arnaud, K.~A.\ 1996, 
Astronomical Data Analysis Software and Systems V, 101, 17 

\bibitem[Begelman(2002)]{begelman02} Begelman, M.C.\ 2002, ApJ, 568,
L97

\bibitem[Begelman(2006)]{begelman06} Begelman, M.C., King, A.R.,
Pringle, J. E.\ 2006, MNRAS, 370, 399	

\bibitem[Bautz et al.(1998)]{bautz98} Bautz, M.~W., et al.\ 1998, \procspie, 3444, 210 

\bibitem[Colbert \& Mushotzky(1999)]{colbert99} Colbert, E.J.M.\ \&
Mushotzky, R.F.\ 1999, ApJ, 519, 89

\bibitem[Dubner et al.(1998)]{dubner98} Dubner, G.M., Holdaway, M.,
Goss, W.M., Mirabel, I.F.\ 1998, AJ, 116, 1842

\bibitem[Dolphin(2000)]{dolphin00} Dolphin, A.E.\ 2000, PASP, 112, 1383

\bibitem[Fabbiano(2006)]{fabbiano06} Fabbiano, G.\ 2006, \araa, 
44, 323 

\bibitem[Fabrika \& Mescheryakov(2001)]{fabrika01} Fabrika, S.\ \&
Mescheryakov, A.\ 2001, IAUS, 205, 268

\bibitem[Green(2004)]{green04} Green, D.A.\ 2004, Bull.\ Astron. Soc.\
India, 32, 335

\bibitem[Kaaret et al.(2001)]{kaaret01} Kaaret, P.\ et al.\ 2001,
MNRAS, 321, L29

\bibitem[Kaaret et al.(2003)]{kaaret03} Kaaret, P., Corbel, S.,
Prestwich, A.H., Zezas, A.\ 2003, Science,  299, 365.

\bibitem[Kaaret et al.(2005)]{kaaret05} Kaaret, P.\ 2005, ApJ, 629, 233

\bibitem[Karachentsev et al.(2002)]{karachentsev02} Karachentsev, I.D.\
et al.\ 2002, A\&A, 385, 21

\bibitem[King et al.(2001)]{king01} King, A.R.\ et al.\ 2001, ApJ, 552,
L109

\bibitem[K\"ording et al.(2002)]{kording02} K\"ording, E. Falcke, H.,
\& Markoff, S.\ 2002, A\&A, 382, L13

\bibitem[Makishima et al.(2000)]{makishima00} Makishima, K.\ et al.\
2000, ApJ, 535, 632

\bibitem[Margon(1984)]{margon84} Margon, B.\ 1984, ARA\&A, 22, 507

\bibitem[Miller et al.(2005)]{miller05} Miller, N.A, Mushotzky, R.F.,
Neff, S.G.\ 2005, ApJ, 623, L109

\bibitem[Remillard \& McClintock(2006)]{remillard06} 
Remillard, R.~A., \& McClintock, J.~E.\ 2006, \araa, 44, 49 

\bibitem[Rodr\'{\i}guez, Mirabel, Mart\'{\i}(1992)]{rodriguez92} Rodr\'{\i}guez,
L.F., Mirabel, I.F., Mart\'{\i}, J.\ 1992, ApJ, 401, L15

\bibitem[Sault \& Killeen(1998)]{sault98} Sault R.J. \& Killeen N.E.B.
1998, The Miriad User's Guide, Sydney: Australia Telescope National
Facility

\bibitem[Schlegel, Finkbeiner, \& Davis(1998)]{schlegel98} Schlegel,
D., Finkbeiner, D., \& Davis, M.\ 1998, ApJ, 500, 525

\bibitem[Soria et al.(2006)]{soria06} Soria, R., Fender, R.P., 
Hannikainen, D.C., Read, A.M., \& Stevens, I.R.\ 2006, MNRAS, 368, 1527

\bibitem[Skrutskie et al.(2006)]{2mass} Skrutskie, R.M.\ et al.\ 2006,
AJ, 131, 1163.

\bibitem[van Paradijs \& McClintock(1995)]{vp95} van Paradijs, J.\ \&
McClintock J.E.\ in X-Ray Binaries, eds.\ W.H.G.\ Lewin,  J.\ van
Paradijs, \& E.P.J. van den Heuvel, (Cambridge Univ.\ Press, 1995),
pp.\ 58-125.

\bibitem[van Speybroeck et al.(1997)]{vanspey97} van Speybroeck, 
L.~P., Jerius, D., Edgar, R.~J., Gaetz, T.~J., Zhao, P., \& Reid, P.~B.\ 
1997, \procspie, 3113, 89 

\bibitem[Weisskopf(1988)]{weisskopf88} Weisskopf, M.~C.\ 1988, 
Space Science Reviews, 47, 47 

\end{thebibliography}
\end{document}